\documentclass[final,5p,times,twocolumn]{elsarticle}

\usepackage{mathtools}
\usepackage{amssymb}
\usepackage{bm}

\usepackage{microtype}

\usepackage{hyperref}
\hypersetup{
	colorlinks=true,
	linkcolor=blue,
	citecolor=blue,
	urlcolor=blue
}

\biboptions{numbers,square,sort&compress}
\journal{Physics Letters B}

\newcommand{\GeV}{\mathrm{GeV}}
\newcommand{\dd}{\mathrm{d}}
\newcommand{\rmbg}{r_m^{\mathrm{bg}}}
\newcommand{\mE}{\mathcal{E}}


\begin{document}
	
	\begin{frontmatter}
		
		\title{Strongly supercooled first-order phase transitions: monopole abundance, entropy production and homogeneity consistency}
		
		\author[inst1]{Chen Zhang\corref{cor1}}
		\ead{zhangchen2@mail.neu.edu.cn}
		
		\author[inst1]{Yue Niu}
		
		\author[inst1,inst2,inst3]{Xin Zhang\corref{cor1}}
		\ead{zhangxin@neu.edu.cn}
		
		\cortext[cor1]{Corresponding authors.}
		
		\affiliation[inst1]{
			organization={Liaoning Key Laboratory of Cosmology and Astrophysics, College of Sciences,
				Northeastern University, No.~3-11, Wenhua Road, Heping District},
			city={Shenyang 110819},
			country={China}
		}
		
		\affiliation[inst2]{
			organization={MOE Key Laboratory of Data Analytics and Optimization for Smart Industry,
				Northeastern University, No.~3-11, Wenhua Road, Heping District},
			city={Shenyang 110819},
			country={China}
		}
		
		\affiliation[inst3]{
			organization={National Frontiers Science Center for Industrial Intelligence and Systems Optimization,
				Northeastern University, No.~3-11, Wenhua Road, Heping District},
			city={Shenyang 110819},
			country={China}
		}
		
		\begin{abstract}
			We investigate whether monopole production in the early universe from a first-order phase transition in a unified gauge theory can be significantly suppressed if the phase transition is subject to strong supercooling but still completes successfully. This is facilitated by adopting a Gaussian nucleation model and assuming the false vacuum energy released is transferred to sectors which behave effectively as radiation in terms of cosmological evolution. We distinguish three criteria of phase-transition completion: the comoving measure completion (CMC) characterized by the eventual vanishing of the false-vacuum fraction (FVF), the physical volume completion (PVC) characterized by the eventual vanishing of the physical volume of false vacuum, and the asymptotic thermodynamic completion (ATC), which refers to the finiteness of total entropy production. Assuming an idealized Friedmann-Lemaître-Robertson-Walker (FLRW) cosmology, it is shown that the three criteria share exactly the same completion boundary, away from which they are completely equivalent. It is further shown that monopole abundance can be arbitrarily suppressed for parameter points approaching the completion boundary from the successful side. Nevertheless, for parameter points too close to the completion boundary, justification of the self-consistency of a homogeneous FLRW cosmology is difficult. We propose the use of the Hubble-scale FVF root-mean-square fluctuation to diagnose the homogeneity consistency issue, and find that arbitrarily strong suppression of monopole abundance relative to a reasonable Kibble estimate is not justifiable in terms of homogeneity consistency.
			These findings also have implications for gravitational wave production in strongly supercooled phase transitions.
		\end{abstract}
		
		\begin{keyword}
			Cosmological phase transition \sep
			Magnetic monopole \sep
			Entropy production
		\end{keyword}
		
	\end{frontmatter}

	\section{Introduction}
	
	Electric charge quantization~\cite{Foot:1990uf,Foot:1992ui} is a notable and intriguing feature of the Standard Model of particle physics. While a consensus on its origin is still lacking, an elegant and compelling explanation is given by the existence of magnetic monopoles~\cite{Preskill:1984gd,Shnir:2005vvi,Rajantie:2012xh,Patrizii:2015uea,Mavromatos:2020gwk,Mitsou:2026zcf}, as Dirac has shown that quantum-mechanical consistency requires that the electric and magnetic charges satisfy the Dirac quantization condition~\cite{Dirac:1931kp}. Moreover, magnetic charges are expected to exist in a consistent theory of quantum gravity~\cite{Witten:2002wb,Polchinski:2003bq,Garcia-Valdecasas:2024cqn}. In grand or partially unified gauge theories, magnetic monopoles are predicted to exist as topological defects~\cite{tHooft:1974kcl,Polyakov:1974ek,Kibble:1976sj,Vilenkin:2000jqa,Weinberg:2012pjx,Lazarides:1980cc,Senoguz:2015lba,Kephart:2017esj,Lazarides:2019xai,King:2020hyd,Lazarides:2021tua,Kephart:2025tik} and be produced via cosmological phase transitions~\cite{Preskill:1979zi,Rajantie:2002dw,Murayama:2009nj,Hindmarsh:2025vxh,Di:2026bnb}. If they were produced before inflation, their abundance would be diluted to an unobservable level. If they were produced during or after inflation, the fact that they have been compatible with astrophysical constraints~\cite{Turner:1982ag,Adams:1993fj,Kobayashi:2023ryr,Zhang:2024mze,Zhang:2026kgb} and escaped all detection efforts~\cite{Price:1986ky,MACRO:2002jdv,Burdin:2014xma,PierreAuger:2016imq,Bai:2021ewf,IceCube:2021eye,Perri:2025qpg,LHAASO:2026ijx} so far means that their abundance must still be sufficiently small. The observation of even a single magnetic monopole may provide a striking confirmation of symmetries at extremely high energies. On the other hand, the failure to observe any magnetic monopoles is by itself also an important piece of information, suggesting the operation of certain mechanisms in the early universe to suppress the monopole abundance.

    For magnetic monopoles produced after inflation, various non-inflationary mechanisms have been proposed in the literature to suppress their abundance~\cite{Langacker:1980kd,Izawa:1982cu,Dvali:1997sa,Bajc:1997ky,Stojkovic:2004hz,Zhang:2023tfv,Zhang:2023zmb,Senjanovic:2025enc,Perri:2025vtn}. In this work, we are interested in the classic proposal~\cite{Guth:1979bh,Einhorn:1980ik,Guth:1981uk,Cook:1981fz} that monopole abundance could be significantly suppressed if they were produced in a strongly supercooled first-order phase transition (FOPT)~\cite{Hindmarsh:2020hop,Wang:2020jrd,Eichhorn:2020upj,Athron:2023xlk}. This is motivated by an apparent tension between a naive Kibble estimate ~\cite{Kibble:1976sj} which seems to disfavor an arbitrarily suppressed monopole abundance, and the observation that a strongly supercooled FOPT may lead to large entropy production. Here one should only consider FOPTs that can complete successfully, and the criteria of phase-transition completion should be determined properly. We also note that strong supercooling may lead to prolongation of the transition process~\cite{Witten:1980ez,Hawking:1981fz,Creminelli:2001th,Randall:2006py,Nardini:2007me,Konstandin:2011dr,Megevand:2016lpr,vonHarling:2017yew,Iso:2017uuu,Ellis:2018mja,Baratella:2018pxi,Fujikura:2019oyi,Brdar:2019qut,DelleRose:2019pgi,Baldes:2021aph,Levi:2022bzt,Sagunski:2023ynd,Agrawal:2025wvf,Hashino:2025fse,Cao:2025jwb,Ning:2026nfs}, while the usual Kibble estimate implicitly assumes a fast transition. An appropriate treatment of the Kibble estimate for prolonged transitions is required to deliver a fair benchmark for comparison of monopole yields.
    
    A further complication of strongly supercooled FOPTs arises from the possible appearance of large bubbles which raises concern about the consistency of the usual assumption that the cosmological expansion is spatially homogeneous and isotropic. The conventional analysis of phase-transition dynamics through the Johnson-Mehl-Avrami-Kolmogorov (JMAK) equation and the fractional extended volume of true bubbles~\cite{Turner:1992tz,Guo:2023gwv} is based on a Friedmann-Lemaître-Robertson-Walker (FLRW) cosmology, and may thus be affected if inhomogeneities caused by large bubbles are significant~\cite{Giombi:2023jqq,Giombi:2025tkv}. Since solving the phase-transition dynamics beyond an FLRW cosmology can be extremely difficult, here we aim at characterizing the degree of inhomogeneities during the phase transition, on which we base our assessment of the consistency of an FLRW treatment of the FOPT.
    
    In this work, the above considerations are investigated in a quantitative manner by adopting a well-motivated Gaussian nucleation model~\cite{Megevand:2016lpr} and assuming the false vacuum energy released is transferred to sectors which have an effective $w=\frac{1}{3}$ equation of state in terms of cosmological expansion, but need not thermalize until very late times. For strongly supercooled FOPTs this is a reasonable simplification with a small set of phase transition parameters capable of capturing the key features of the problem. We distinguish three criteria of phase-transition completion which are naturally introduced according to the asymptotic behavior of the false vacuum fraction (FVF), the physical volume of false vacuum, and a dedicated variable related to thermodynamic completion of the transition. We demonstrate that the three criteria of phase-transition completion share the same completion boundary in the parameter space, which is also well separated from the boundary of homogeneity consistency. While the relative monopole yield defined with respect to an appropriately chosen Kibble estimate can be arbitrarily suppressed near the boundary of phase-transition completion, on the boundary of homogeneity consistency arbitrarily strong suppression of the relative monopole yield cannot be achieved. In the parameter region between the two boundaries, a reliable estimate of entropy production during FOPTs is still lacking, which has important implications not only for the eventual abundance of topological defects, but also for other interesting observables such as phase-transition gravitational waves.
    
    The paper is organized as follows. Sec.~\ref{sec:model} introduces the Gaussian nucleation model and key assumptions of false vacuum energy transfer employed to model strongly supercooled FOPTs, including the associated phase transition dynamics and evolution equations. Sec.~\ref{sec:criteria} then discusses three criteria of phase transition completion and their relations with implications for suppression of the monopole yield near the completion boundary. Sec.~\ref{sec:homogeneity} suggests the use of the Hubble-scale FVF root-mean-square (RMS) fluctuation to diagnose the homogeneity consistency issue.
    In Sec.~\ref{sec:results} numerical results concerning the evolution of characteristic quantities during the phase transition are analyzed; an overview of the boundary of phase transition completion and the boundary of homogeneity consistency is also presented, with an emphasis on relative monopole yields along the latter. Finally in Sec.~\ref{sec:dnc}, a summary with implications for future directions is presented. We adopt natural units with $c = \hbar = k_B = 1$ throughout the paper.

	\section{Modeling of strongly supercooled first-order phase transitions}\label{sec:model}
	
	The bubble nucleation rate of a first-order cosmological phase transition is determined by the thermal or quantum effective action of the underlying model of particle physics. At this moment, very little about the physics beyond the Standard Model is known, and the mapping from new physics parameters to the thermal or quantum effective actions suffers from large uncertainties~\cite{Croon:2020cgk,Arunasalam:2021zrs,Athron:2022jyi,Bittar:2025lcr}. Therefore, in this work we start from an effective parametrization of the nucleation rate. For strongly supercooled transitions, a natural and well-motivated parametrization is given by the Gaussian form~\cite{Megevand:2016lpr,Eichhorn:2020upj,Athron:2023aqe}
	\begin{align}
		\Gamma(t) = \Gamma_m \exp\left[-\alpha^2(t-t_m)^2\right]. \label{eq:gaussian}
	\end{align}
	Here $\Gamma(t)$ is the bubble nucleation rate as a function of the cosmic time $t$, which reaches its maximum $\Gamma_m$ at $t=t_m$; $\alpha^{-1}$ characterizes the width of the Gaussian peak and has the dimension of time. This Gaussian modeling should be understood as an approximation of the nucleation rate computed from the thermal effective action of the underlying particle physics model, expected to capture the behavior of the nucleation rate around its maximum. In many cases bubble expansion in strongly supercooled FOPTs proceeds in the detonation regime~\cite{Espinosa:2010hh}, and the reheating effect in the unconverted false vacuum region is expected to be small~\cite{Megevand:2016lpr,Athron:2022mmm}. Therefore, the Gaussian nucleation model is expected to be a good approximation in such cases. Mild homogeneous reheating can also be absorbed into the parametrization itself. On the other hand, we note that for FOPTs with extreme supercooling, at low temperatures quantum tunneling may dominate over thermal transitions. In such cases, eq.~\eqref{eq:gaussian} fails to capture the contribution from quantum tunneling. For simplicity, this model-dependent quantum-tunneling contribution is neglected in this work and left for future studies.
	
	When eq.~\eqref{eq:gaussian} is adopted, it is implicitly assumed that the temperature in the false vacuum region, as well as the cosmological scale factor $a(t)$, is spatially homogeneous and isotropic. At time $t$, the comoving radius of a bubble nucleated at $t'$ can be expressed as
	\begin{align}
		r(t,t')=v_w[\eta(t)-\eta(t')],\qquad \eta(t) = \eta_i + \int_{t_i}^{t} \frac{\dd t'}{a(t')}. \label{eq:radius}
	\end{align}
	Here $v_w$ is the bubble wall velocity which we assume to be constant for simplicity, and $\eta$ is the conformal time with an initial value $\eta_i$ corresponding to some convenient initial cosmic time $t_i$ when the nucleation rate is negligible. The phase-transition dynamics is characterized by the JMAK equation which delivers the FVF $f_F$~\cite{Turner:1992tz,Megevand:2016lpr}
	\begin{align}
		f_F(t)=e^{-I(t)},\qquad I(t) = \frac{4\pi}{3} \int_{t_i}^{t} \dd t'\, \Gamma(t') a^3(t') r^3(t,t'), \label{eq:JMAK}
	\end{align}
	with $I(t)$ being the fractional extended volume of true bubbles. Regarding the energy contents of the universe during the phase transition, we assume that initially the universe is radiation-dominated before the phase transition, and during the phase transition, the false vacuum energy released is converted to sectors which behave effectively as radiation in terms of cosmological expansion. We only require these sectors thermalize at late times, but thermal equilibrium during the phase transition may not be satisfied. For strongly supercooled FOPTs, such sectors contain all forms of highly relativistic objects such as bubble walls, non-thermal and thermal particle distributions fueled by continuous conversion of the false vacuum energy, and will be called radiation-like sectors in the following. The evolution of the scale factor is then determined by the Friedmann equation\footnote{Vacuum energy density of the true vacuum is assumed to be zero.}
	\begin{align}
		H^2 = \frac{\rho_R + \Delta V_0 f_F}{3M_{\mathrm{Pl}}^2}. \label{eq:Friedmann}
	\end{align}
	Here $H=\dot{a}/a$ is the Hubble parameter ($\dot{a}\equiv\frac{\dd a}{\dd t}$), $M_{\mathrm{Pl}}\simeq 2.4\times 10^{18}\,\GeV$ is the reduced Planck mass, $\Delta V_0$ is the energy density difference between the false vacuum and the true vacuum at zero temperature, and $\rho_R$ is the cosmologically-averaged energy density of the radiation-like sectors which can belong to unconverted or converted regions, and also bubble walls. The evolution of $\rho_R$ is given by~\cite{Liu:2022lvz,Gouttenoire:2023naa,Lewicki:2023ioy,Lewicki:2024sfw,Ghosh:2026gmp}
	\begin{align}
		\dot{\rho}_R + 4H\rho_R = -\Delta V_0 \dot{f}_F. \label{eq:VRB}
	\end{align}

    Eqs.~\eqref{eq:gaussian}--\eqref{eq:VRB} provide a closed system of equations simultaneously tracking the progress of the phase transition and the cosmological expansion. They can be solved numerically once the initial conditions are specified. The mutual influence of the phase-transition dynamics and the cosmological expansion is incorporated into the equations in a self-consistent manner, as long as the FLRW description remains valid during the phase transition.
    
    For convenience, we introduce the vacuum Hubble scale $H_V = \sqrt{\frac{\Delta V_0}{3M_{\mathrm{Pl}}^2}}$, and dimensionless parameters
    \begin{align}
    	\lambda_V = \frac{(\Delta V_0)^{1/4}}{M_{\mathrm{Pl}}},\quad r_m^{\mathrm{bg}} = \frac{\rho_R^{\mathrm{bg}}(t_m)}{\Delta V_0},\quad q = \frac{\alpha}{H_V}, \quad \gamma = \frac{\Gamma_m}{H_V^4}.
    \end{align}
    Here $\rho_R^{\mathrm{bg}}(t_m)$ is the radiation energy density evaluated at $t_m$ in the counterfactual evolution without the phase transition, assuming the same initial condition as the actual evolution in the presence of the phase transition. $r_m^{\mathrm{bg}}$ is then the dimensionless parameter specifying the initial condition of the problem. $\gamma$ and $q$ are the dimensionless parameters which control the nucleation amplitude and the Gaussian width. While $\lambda_V$ controls the overall scale of the problem, quite a few features of the phase transition are scale-independent and only depend on the dimensionless set $(\rmbg,q,\gamma,v_w)$.
	
    \section{Criteria of phase-transition completion and suppression of the monopole yield}\label{sec:criteria}
    
    The competition between the conversion and the expansion of the false vacuum raises the issue of phase-transition completion~\cite{Guth:1982pn,Turner:1992tz}. In the literature, completion criteria are usually based on whether the behavior of the FVF and the physical volume of the false vacuum can reach certain prescribed thresholds~\cite{Megevand:2016lpr,Athron:2022mmm}. For instance, one may check whether the physical volume of the false vacuum is decreasing when the FVF reaches a prescribed small number such as $0.01$. Such criteria are usually sufficient for practical purposes, but they are not equivalent to the expected asymptotic notion of phase-transition completion. In this section, we introduce three criteria of phase-transition completion defined in terms of the asymptotic, limiting behavior of the phase-transition dynamics.
    
    The first completion criterion is termed comoving measure completion (CMC), defined as
    \begin{align}
	    \mathrm{CMC}: \quad \lim_{t \to \infty} f_F(t) = 0.
    \end{align}
    The second completion criterion is termed physical volume completion (PVC), defined as
    \begin{align}
    	\mathrm{PVC}: \quad \lim_{t \to \infty} a^3(t)f_F(t) = 0.
    \end{align}
    These two criteria fail to account for FOPTs which lead to eternal inflation or permanent false vacuum remnants but are still compatible with observations. Moreover, although PVC also deals with the physical volume of the false vacuum, it is technically different from the more common requirement that the physical volume of the false vacuum is decreasing at certain benchmark moments during the transition. Nevertheless, the CMC and PVC introduced above are natural, universal criteria for phase-transition completion in the sense that they do not depend on any benchmark moments chosen artificially during the transition.
    
    The third completion criterion is termed asymptotic thermodynamic completion (ATC), introduced to allow for finite entropy production during the phase transition. Technically, it is convenient to define the comoving radiation-energy budget
    \begin{align}
    	\mathcal{E}=a^4 \rho_R.
    \end{align}
    ATC is then defined through the asymptotic condition
    \begin{align}
 	    \mathrm{ATC}: \quad \lim_{t \to \infty} \mathcal{E}(t) < \infty. \label{eq:ATC}
    \end{align}   
    The importance of the comoving radiation-energy budget is embodied in the fact that if the radiation-like sectors are all in thermal equilibrium, then the comoving entropy density $S_c$ is directly proportional to $\mathcal{E}^{3/4}$:
    \begin{align}
    	S_c = C_s \mathcal{E}^{3/4},\qquad C_s\equiv\frac{2\pi^2}{45} g_*^s \left(\frac{30}{\pi^2 g_*^\rho}\right)^{3/4}, \label{eq:Sc}
    \end{align}
    with $g_*^\rho$ and $g_*^s$ being the relativistic effective degrees of freedom in terms of energy and entropy, respectively. Therefore, for constant $C_s$, conservation of $\mathcal{E}$ is equivalent to entropy conservation when thermal equilibrium can be used. For strongly supercooled FOPTs considered here, thermal equilibrium is assumed to be valid only at the beginning and the end of the transition (if the transition completes successfully), while intermediate states may feature highly non-thermal particle distributions and scalar-field motions. Therefore, eq.~\eqref{eq:ATC} employs $\mathcal{E}(t)$ rather than $S_c$ to define ATC in an asymptotic manner. 
    
    In this work, we are concerned with FOPTs in grand or partially unified gauge theories which lead to monopole production due to symmetry breaking patterns with a nontrivial second homotopy group. For FOPTs with successful ATC, the final comoving entropy at the end of the transition is finite, and thus the monopole yield $Y_M$ can be consistently defined as
    \begin{align}
    	Y_M=\lim_{t \to \infty}\frac{N_M}{S_c}, \label{eq:YM}
    \end{align}
    where $N_M$ is the comoving number density of monopoles. The right-hand side of eq.~\eqref{eq:YM} is supposed to be evaluated at a moment late enough so that the transition is effectively completed, but also early enough so that the post-transition monopole annihilation is not significant yet.
    
    In principle, the three completion criteria can be defined for cosmological FOPTs with arbitrary nucleation rates. Nevertheless, the assumption of Gaussian nucleation rates is both simple and well-motivated, and allows us to have both analytical and numerical understanding of the relationship between the three completion criteria. For simplicity, we fix $v_w=1$ so that for scale-independent features, the relevant phase-transition parameter space is spanned by $(\rmbg,q,\gamma)$.
    Each of the three completion criteria should define a boundary surface in this parameter space, which separates FOPTs that complete successfully or fail to complete. The natural questions are then the relationship between the three boundary surfaces, completion status for points exactly on the boundary, and the behavior of the monopole yield near the boundary.
    
    From the definitions, one obvious relation is that PVC implies CMC, since the scale factor is always increasing. As to the remaining relations, within the Gaussian nucleation model and energy-transfer assumption adopted in this work, it is possible to prove the following propositions:
    \begin{enumerate}
    	\item ATC and PVC are equivalent.
    	\item Away from the completion boundary, CMC implies PVC.
    	\item CMC, PVC and ATC share the same completion boundary, on which CMC succeeds while PVC and ATC fail.
    	\item For fixed $\lambda_V$, the monopole yield can be arbitrarily suppressed by approaching the completion boundary from the successful side.
    \end{enumerate}
    A sketch of the proof of the first three propositions is presented in the Supplementary Material. Once the first three propositions are proved, the path leading to the fourth proposition then becomes clear. The basic logic is reduction to absurdity. The fourth proposition basically states that the upper limit of the entropy production (or equivalently, the comoving radiation-energy budget) of the FOPT reaches infinity as one approaches the completion boundary from the successful side. If this were false, that would mean there existed a way to approach the completion boundary with the entropy production being constrained by a finite upper limit. Then the limiting point on the completion boundary would also have a finite entropy production, implying the success of ATC on the boundary, in contradiction to the third proposition.
    
    Let us note two caveats in interpreting the concepts and propositions about these completion criteria. First, these three criteria and the associated propositions are all based on the assumption that an FLRW description of the cosmological expansion remains consistent during the phase transition. In a realistic analysis this consistency needs to be checked, as discussed in Sec.~\ref{sec:homogeneity}. Barring this homogeneity consistency issue, the concepts and propositions studied in this section should be understood as an ideal mathematical modeling of the phase-transition dynamics. We expect that the idealized modeling may still capture some features of the realistic phase-transition dynamics in the regime in which inhomogeneities have a large impact, and therefore the idealized modeling may still be of great value given the lack of reliable simulations in such regimes. Second, bubble nucleation is a stochastic process, while the three criteria and eqs.~\eqref{eq:gaussian}--\eqref{eq:VRB} are based on quantities defined in the sense of ensemble average. Therefore all the results obtained here should also be understood in the sense of ensemble average rather than any specific realization of the nucleation process.
    
    \section{Assessment of the homogeneity consistency}\label{sec:homogeneity}
    
    While bubble nucleation is stochastic, if a large number of bubbles are nucleated in one Hubble volume, then the assumption of an FLRW cosmology is expected to be valid on Hubble scales. The inhomogeneities are expected to be significant if in one Hubble volume, only a small number of large bubbles with sizes comparable to the Hubble scale are present during some stage of the FOPT. In such cases, fluctuations of the Hubble-scale FVF are also expected to be significant, which we adopt as an indicator of the degree of the inhomogeneity. To begin with, let the FLRW metric be $\dd s^2 = -\dd t^2 + a^2(t)\dd\boldsymbol{\xi}^2$, and we introduce the dimensionless comoving spatial coordinate $\mathbf{r}=H_V\boldsymbol{\xi}$ and the dimensionless time coordinate $x=H_V(t-t_m)$. The false-vacuum indicator field~\cite{Wang:2025ooq} is defined as
    \begin{align}
    	\chi_F(\mathbf{r},x) =
    	\begin{cases}
    		1, & \mathbf{r} \text{ is in the false-vacuum phase at } x,\\
    		0, & \mathbf{r} \text{ has already transitioned at } x.
    	\end{cases}
    \end{align}
    We then introduce the spherical Hubble smoothing region around a spatial center $\mathbf{X}$ at time $x$:
    \begin{align}
    	D_H(\mathbf{X},x) = \left\{ \mathbf{r} \in \mathbb{R}^3 : |\mathbf{r}-\mathbf{X}| \leq R_H(x) \right\},
    \end{align}
    where $R_H(x)\equiv\frac{H_V}{a(x)H(x)}$ is the dimensionless comoving radius of the smoothing sphere. The Hubble-scale FVF is then defined as
    \begin{align}
    	F_H(\mathbf{X},x) = \frac{1}{V_H(x)} \int_{D_H(\mathbf{X},x)} \dd^3r\, \chi_F(\mathbf{r},x), \label{eq:FH}
    \end{align}
    where $V_H(x)\equiv\frac{4\pi}{3}R_H^3(x)$. Its RMS fluctuation is given by
    \begin{align}
    	\sigma_{F,H}(x) = \sqrt{\sigma^2_{F,H}(x)},\quad\sigma^2_{F,H}(x) \equiv \mathrm{Var}_{\mathrm{nuc}}\left[F_H(\mathbf{X},x)\right]. \label{eq:sigmaFH}
    \end{align}
    Here $\mathrm{Var}_{\mathrm{nuc}}$ means the variance is evaluated by considering ensemble average of different nucleation realizations, which has the consequence that $\sigma^2_{F,H}$ becomes independent of $\mathbf{X}$. Nevertheless, $\sigma^2_{F,H}$ still depends on time. Thus we introduce a phase-homogeneity indicator at the trajectory level:
    \begin{align}
    	\Xi_F \equiv \max_{x} \sigma_{F,H}(x).
    \end{align}
    For practical purposes, we adopt $\Xi_F=\Xi_{F,\rm crit}\equiv 0.2$ as the critical threshold for the boundary of homogeneity consistency. Roughly speaking, if the maximum of $\sigma_{F,H}$ is reached when $f_F=0.5$, this suggests the $1\sigma$ fluctuation range of $F_H$ is $0.5\pm 0.2$.
    
    For the Gaussian nucleation model considered in this work, let $y$ be the dimensionless nucleation time of some bubble, and define 
    \begin{align}
    	S(y)=e^{-q^2 y^2+3u(y)},\qquad u(y)\equiv\ln a(y),
    \end{align}
    which is the function that always appears in the nucleation intensity measure. $R(x,y)=v_w\int_y^x\dd x'e^{-u(x')}$ will be used to denote the comoving radius of the bubble at time $x$. It is possible to derive a simple integration formula to evaluate $\sigma^2_{F,H}(x)$:
    \begin{align}
    	\sigma^2_{F,H}(x) = f_F^2(x) \int_0^2 \dd z\, K(z) \left[e^{J(zR_H(x),x)} - 1\right]. \label{eq:sigmaFH2}
    \end{align}
    Here $K(z)$ is given by
    \begin{align}
    	K(z) = \frac{3}{16}z^2(4+z)(2-z)^2, \qquad 0 \leq z \leq 2, \label{eq:kernel}
    \end{align}
    and the overlap exponent function $J(s,x)$ is defined as
    \begin{align}
    	J(s,x) = \gamma \int_{x_i}^{x} \dd\,yS(y)V_{\cap}[R(x,y),s], \label{eq:overlap}
    \end{align}
    where $V_{\cap}[R(x,y),s]$ is the overlapping volume of two spheres of comoving radius $R(x,y)$ whose centers are separated by $s$
    \begin{align}
    	V_{\cap}(R,s) =
    	\begin{cases}
    		\displaystyle \frac{\pi}{12}(4R+s)(2R-s)^2, & 0 \leq s < 2R,\\
    		0, & s \geq 2R.
    	\end{cases} \label{eq:overlapV}
    \end{align}
    The derivation of eq.~\eqref{eq:sigmaFH2} is given in ~\ref{sec:app}.

    \section{Numerical results}\label{sec:results}
    
\begin{figure}[t]
		\centering
		\includegraphics[width=\linewidth]{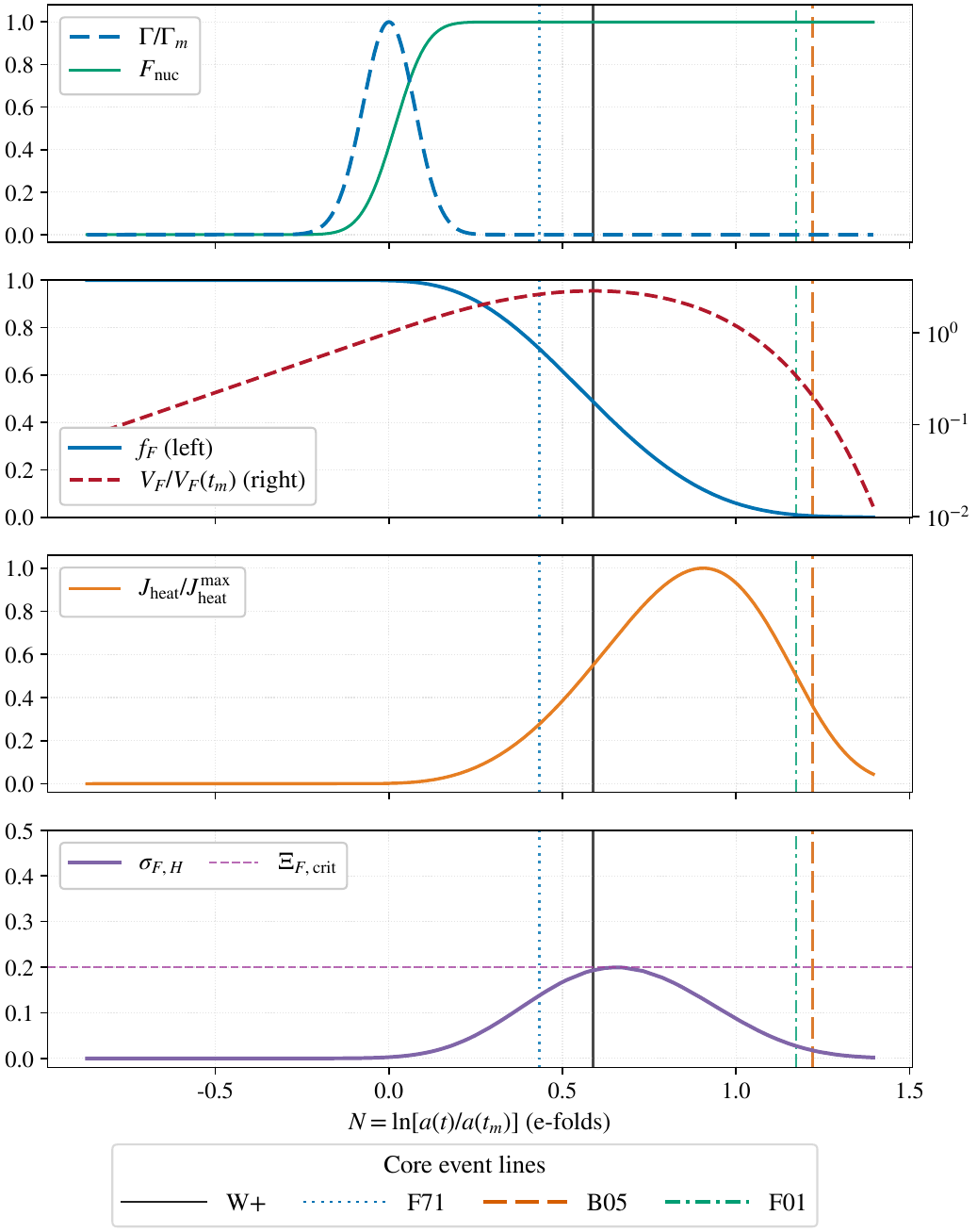}
		\caption{
			Evolution of characteristic quantities as a function of the number of e-folds associated with the cosmic expansion. The model parameters are $\rmbg=0.1,q=10,\gamma=10.5898,v_w=1$. From top to bottom: Panel 1 shows the normalized attempted nucleation rate $\Gamma/\Gamma_m$ and the actual nucleation fraction $F_{\rm nuc}=C_0(x)/C_{0,\infty}$. Panel 2 shows the FVF (left axis) and the physical volume of the false vacuum divided by its value at the Gaussian peak. Panel 3 shows the normalized ``heat injection'' variable $J_{\rm heat}/J_{\rm heat}^{\rm max}$. Panel 4 shows the Hubble-scale false-vacuum-fraction root-mean-square fluctuation $\sigma_{F,H}$ and the horizontal reference line at $\Xi_{F,\rm crit}$. Four core event lines are displayed across all panels: $\rm W+$ corresponds to the moment when the physical volume of the false vacuum starts to decrease and never increases again, $\rm F71$ and $\rm F01$ correspond to $f_F=0.71$ and $f_F=0.01$ respectively, and $\rm B05$ corresponds to the moment when only $5\%$ of the total ``heat injection'' remains to be completed.
		}
		\label{fig:one}
	\end{figure}

    In this section, eqs.~\eqref{eq:gaussian}--\eqref{eq:VRB} are numerically solved for certain regions of the parameter space, while $\sigma^2_{F,H}(x)$ and $\Xi_F$ are evaluated to determine the homogeneity consistency of the calculation. First, we choose the benchmark parameter point
    \begin{align}
    	\rmbg=0.1,\,q=10,\,\gamma=10.5898,\,v_w=1,
    \end{align}
    which is close to the boundary of homogeneity consistency (i.e. $\Xi_F\simeq 0.2$), to illustrate the progress of the phase transition. The results are displayed in Fig.~\ref{fig:one}, in which $N\equiv\ln\frac{a(t)}{a(t_m)}$ is chosen to be the horizontal axis for all panels, while the vertical axes in these panels contain quantities reflecting multiple aspects of the phase transition: attempted and actual nucleation (Panel 1), FVF and the physical volume of the false vacuum (Panel 2), ``heat injection'' (Panel 3), and the phase-homogeneity indicator (Panel 4). Here the ``heat injection'' variable is defined as $J_{\rm heat}\equiv\frac{\dd E}{\dd N}$ which measures how fast the comoving radiation-energy budget increases with $N$. Prolongation of the transition at this parameter point is manifest, as can be seen from the displacements among features of various curves shown in the figure. It is remarkable that the peak of ``heat injection'' is delayed by around 1 e-fold compared to the peak of Gaussian nucleation. A larger delay of the ``heat injection'' can be obtained by decreasing $\gamma$ while keeping other parameters fixed, which may however lead to larger $\Xi_F$ and thus the homogeneity consistency would be difficult to justify.

	
    Next, we investigate the suppression of monopole yields. Define
    \begin{align}
	    C_0(x) = \int_{x_i}^{x} \dd y\, S(y) f_F(y),
    \end{align}
    then the comoving number density of the actually nucleated bubbles is
    \begin{align}
	    N_b(x)=H_V^3\gamma C_0(x).
    \end{align}
    The comoving number density of monopoles $N_M$ is assumed to be proportional to $N_b(\infty)$:
    \begin{align}
	    N_M=\kappa_M N_b(\infty)=\kappa_M H_V^3\gamma C_{0,\infty},\qquad C_{0,\infty}\equiv C_0(\infty), \label{eq:NM}
    \end{align}
    where $\kappa_M$ is an $\mathcal{O}(1)$ constant computed from lattice simulations~\cite{Di:2026bnb}. The absolute monopole yield $Y_M^{\mathrm{abs}}$ is obtained by combining eqs.~\eqref{eq:Sc}~\eqref{eq:YM}~\eqref{eq:NM}:
    \begin{align}
	    Y_M^{\rm abs} = \frac{\kappa_M \gamma C_{0,\infty} \lambda_V^3}{3^{3/2} C_s E_\infty^{3/4}},\qquad E_\infty\equiv\frac{\mathcal{E}(\infty)}{\Delta V_0}.
    \end{align}
    Since $E_\infty$ is independent of $\lambda_V$, we have $Y_M^{\rm abs}\propto\lambda_V^3$. It is helpful to introduce a relative monopole yield
    \begin{align}
    	\hat{Y}_M\equiv\frac{Y_M^{\rm abs}}{Y_{M,K}},
    \end{align}
    where $Y_{M,K}$ is the Kibble estimate obtained by considering a hypothetical fast FOPT occurring in radiation domination at the same critical temperature $T_c$ as the actual supercooled FOPT. With the corresponding Hubble rate and entropy density at $T_c$ given by
    \begin{align}
    	H_c^{\mathrm{RD}} = \sqrt{\frac{\pi^2 g_*^\rho}{90}} \frac{T_c^2}{M_{\mathrm{Pl}}},\qquad s_c^{\mathrm{RD}} = \frac{2\pi^2}{45} g_*^s T_c^3,
    \end{align}
    the physical number density of monopoles in the Kibble benchmark at the end of the transition can be written as
    \begin{align}
    	n_{M,K}=\frac{3}{4\pi}\kappa_K(H_c^{\mathrm{RD}})^3,
    \end{align}
    where $\kappa_K\sim\mathcal{O}(1)$ is the number of monopoles produced for each sphere of physical radius $(H_c^{\mathrm{RD}})^{-1}$. Let $T_c=\chi_c(\Delta V_0)^{1/4}$, the monopole yield in the Kibble estimate is found to be
    \begin{align}
    	Y_{M,K}=\frac{3}{4\pi}\kappa_K C_y \chi_c^3\lambda_V^3,\qquad C_y\equiv\frac{45}{2\pi^2 g_*^s} \left(\frac{\pi^2 g_*^\rho}{90}\right)^{3/2}.
    \end{align}
    Finally, the relative monopole yield $\hat{Y}_M$ can be computed as
    \begin{align}
    	\hat{Y}_M = \frac{4\pi}{3}\frac{\kappa_M}{\kappa_K} \gamma C_{0,\infty} E_\infty^{-3/4} \left(\frac{30}{\pi^2 g_*^\rho}\right)^{3/4} \chi_c^{-3}.
    \end{align}
    
\begin{figure}[t]
		\centering
		\includegraphics[width=\linewidth]{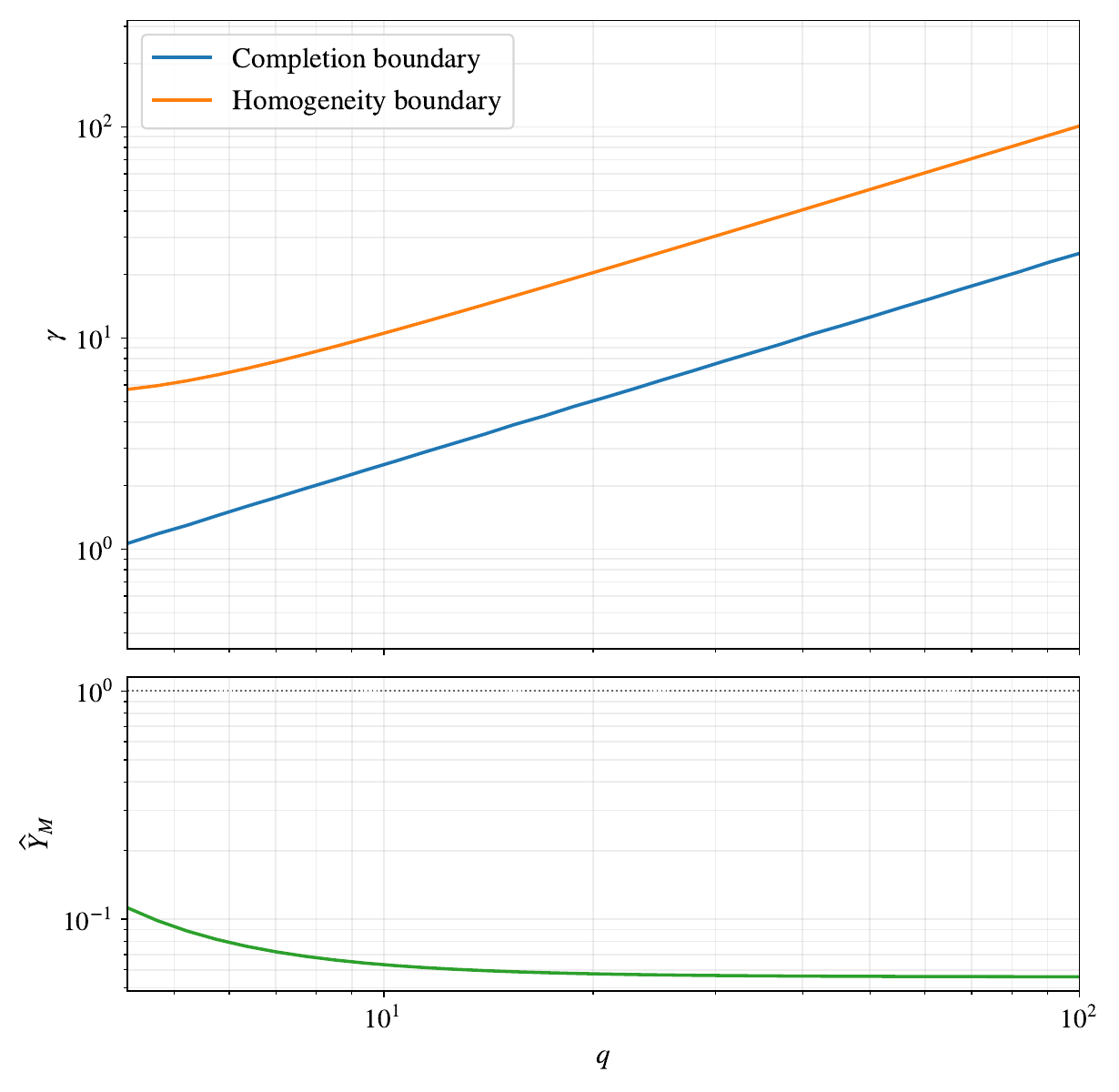}
		\caption{
			Upper panel: the boundary of phase transition completion and the boundary of homogeneity consistency in the $q-\gamma$ plane. Lower panel: relative monopole yield as a function of $q$ along the boundary of homogeneity consistency. In both panels $r_{\rm m}^{\rm bg}=0.1$ and $v_w=1$ are adopted.
		}
		\label{fig:two}
	\end{figure}

    In the upper panel of Fig.~\ref{fig:two}, the boundary of phase-transition completion and the boundary of homogeneity consistency are plotted in the $q-\gamma$ plane, for fixed $\rmbg=0.1$ and $v_w=1$. Numerical scans confirm that the three completion criteria share the same boundary to the available numerical precision. The relative monopole yield $\hat{Y}_M$ along the boundary of homogeneity consistency is plotted in the lower panel of Fig.~\ref{fig:two} as a function of $q$, assuming the same set of parameters and $\kappa_M=\kappa_K=\chi_c=1,g_*^\rho=106.75$. We find that $\hat{Y}_M$ is at the level of a few percent or higher, implying that an arbitrary suppression of $\hat{Y}_M$ cannot be obtained as long as the homogeneity consistency is satisfied. On the other hand, close to the completion boundary (but still from the successful side), the existence of parameter points with strongly suppressed $\hat{Y}_M$ has been confirmed numerically. For instance, $\rmbg=0.1,q=10,\gamma=2.514076,v_w=1$ leads to $\hat{Y}_M\simeq 4.4\times 10^{-14}$, and the peak of ``heat injection'' is found to be delayed by about 17 e-folds compared to the Gaussian nucleation peak. However, for this point $\Xi_F=0.3723$ which definitely exceeds the critical threshold $0.2$, so the homogeneity consistency is not justified. In the numerical exploration, we have varied $\rmbg$ by many orders of magnitude, from $\rmbg=100$ to $\rmbg=10^{-20}$. In all cases satisfying the homogeneity consistency requirement, $\hat{Y}_M$ is found to be at least at the level of a few percent, implying the absence of strong, arbitrary suppression of the monopole yield relative to the Kibble estimate. In particular, for small $\rmbg$ such as $\rmbg=10^{-20}$ which literally corresponds to deep supercooling, while one may obtain a very large ratio of the final to initial entropy and simultaneously satisfy the homogeneity consistency, $\hat{Y}_M$ is still at the level of a few percent. In such cases, nucleation and ``heat injection'' are both delayed relative to $T=T_c$, but the peak of ``heat injection'' is only mildly delayed with respect to the Gaussian nucleation peak by about 1 e-fold, in contrast to the many e-folds of delay found in the near-boundary cases.
	
	\section{Discussion and conclusions}\label{sec:dnc}
	
	In this work, motivated by a potential solution of the monopole problem through strongly supercooled FOPTs, we adopt a Gaussian nucleation model with appropriate vacuum-to-radiation energy-transfer assumptions which enable us to investigate the problem both analytically and numerically. The equations of phase-transition dynamics and cosmological evolution are analyzed in a self-consistent manner which properly takes into account backreaction in both directions. We propose three phase-transition completion criteria (CMC, PVC, ATC) which are defined in terms of asymptotic limiting behavior of the transition to get rid of dependence on some arbitrarily chosen threshold. Intriguingly, in the simple Gaussian nucleation and vacuum-to-radiation energy-transfer model employed in this work, the three completion criteria share the same completion boundary, while exactly on the boundary CMC succeeds but PVC and ATC fail. This has the important consequence that the monopole yield can be arbitrarily suppressed as one approaches the completion boundary from the successful side.
	
	The above results concerning CMC, PVC, ATC and their relations apply to an idealized mathematical model in which the FLRW cosmology is assumed to be valid. In a realistic universe one must examine the consistency in using an FLRW metric and the usual equations of phase-transition dynamics with a universal scale factor. In this regard, we propose the use of the Hubble-scale FVF RMS fluctuation as a phase-inhomogeneity indicator and present compact formulae for its evaluation. We note however it only reflects phase-inhomogeneity, which can be different from inhomogeneities associated with spatial curvature or in the energy-momentum tensor that enters Einstein's equations~\cite{Franciolini:2025ztf,Wang:2026zvz,Li:2026hef}. Therefore, $\Xi_F>0.2$ should be used as a practical reminder of the homogeneity consistency concern, rather than be interpreted as a final verdict on homogeneity failure\footnote{One advantage of using $\Xi_F$ lies in its computational simplicity: cosmological perturbation and numerical simulation are not needed.}.
	
	In the numerical analysis, it is found that the relative monopole yield can only be suppressed to the level of a few percent, but not much lower when the homogeneity consistency is satisfied. Barring the homogeneity consistency issue, strong suppression of the relative monopole yield is confirmed near the completion boundary, which also corresponds to large absolute entropy production. In the model adopted here, the three most relevant phase-transition parameters are $\rmbg,q,\gamma$. While the usual notion of supercooling corresponds most directly to a small $\rmbg$, the boundary of phase-transition completion is found to be reached most easily in the $q-\gamma$ plane, for a fixed $\rmbg$. If the large entropy production near the completion boundary is confirmed in future studies which properly solve the homogeneity issue, it would be important not only for the monopole problem, but also for prediction of phase-transition gravitational waves~\cite{Kobakhidze:2017mru,Ellis:2019oqb,Ellis:2020nnr,Lewicki:2020jiv,Lewicki:2022pdb,Kierkla:2023von,Athron:2023mer,Athron:2023rfq,Conaci:2024tlc,Lewicki:2025hxg,Borah:2026nkw}. The common belief that phase transitions with stronger supercooling feature stronger gravitational-wave signals needs to be reviewed if large entropy production dilutes gravitational waves from the same FOPT. The conclusion could also depend on the direction of variation in the phase-transition parameter space.
	
	Finally, although our analysis is carried out for magnetic monopoles, similar analyses can also be carried out for hidden-sector monopoles~\cite{Baek:2013dwa,Khoze:2014woa,Kawasaki:2015lpf,Nomura:2015xil,Graesser:2020hiv,Yang:2022quy,Zhou:2025zzz,Brummer:2025inh,Liveoak:2026muy}, or other topological defects such as domain walls and cosmic strings.

	
	\section*{Acknowledgements}
	
	This work was supported by the National Natural Science Foundation of China Grants (No. 12533001, No. 12575049 and No. 12473001), the National SKA Program of China Grants (No. 2022SKA0110200 and No. 2022SKA0110203), the China Manned Space Program (Grant No. CMS-CSST-2025-A02), and the 111 Project (Grant No. B16009). ChatGPT (GPT-6) was used to assist with code development and analytical derivation. The authors take full responsibility for the content of this work.

	
    \appendix
    \section{Derivation of the Hubble-scale FVF variance} \label{sec:app}

    For completeness, we outline the derivation of eq.~\eqref{eq:sigmaFH2}. Consider two spatial points $\mathbf{r}_1,\mathbf{r}_2$ in comoving coordinates with the relative displacement $\mathbf{s}=\mathbf{r}_2-\mathbf{r}_1$. At a fixed time $x$, statistical homogeneity and isotropy imply that the two-point covariance of the false-vacuum indicator field depends only on the comoving separation $s\equiv|\mathbf{s}|$. We define
    \begin{align}
	C_F(s,x)\equiv\left\langle	\chi_F(\mathbf{r}_1,x)\chi_F(\mathbf{r}_2,x)\right\rangle_{\rm nuc}	-f_F^2(x),
    \end{align}
    where $\langle...\rangle_{\rm nuc}$ denotes an ensemble average with respect to nucleation realizations. Using eqs.~\eqref{eq:FH}~\eqref{eq:sigmaFH}, the variance of the Hubble-scale FVF can then be expressed as
    \begin{align}
	    \sigma_{F,H}^2(x)=\frac{1}{V_H^2(x)}\int_{D_H}\dd^3r_1\int_{D_H}\dd^3r_2\, C_F(|\mathbf{r}_1-\mathbf{r}_2|,x).
	\label{eq:app_variance_covariance}
    \end{align}

    The definition of the indicator field $\chi_F$ implies that $\langle\chi_F(\mathbf{r}_1,x)\chi_F(\mathbf{r}_2,x)\rangle_{\rm nuc}=P_{FF}(s,x)$, with $P_{FF}(s,x)$ being the joint probability that both points remain in the false-vacuum phase.
    In the Gaussian nucleation model, the dimensionless nucleation intensity measure is $\dd \mu=\gamma S(y)\,\dd y\,\dd^3b$, where $\mathbf{b}$ denotes the dimensionless comoving position of a bubble center. For a given nucleation time $y$, the centers of bubbles that can cover either of the two points $\mathbf{r}_1,\mathbf{r}_2$ by time $x$ form the union of two spheres of radius $R(x,y)$, whose centers are separated by $s$. Let $\mathcal{P}_1$ and $\mathcal{P}_2$ denote the corresponding exclusion regions in nucleation spacetime. Their intersection has intensity measure $J(s,x)$, as defined in eq.~\eqref{eq:overlap}. Since the void probability of each individual exclusion region is $f_F(x)$, the Poisson zero-event probability and the inclusion-exclusion relation give~\cite{Megevand:2020klf}
	\begin{align}
		P_{FF}(s,x) & =\exp\left[-\mu(\mathcal{P}_1\cup\mathcal{P}_2)\right] \nonumber \\
		& = \exp\left[2\ln f_F(x)+J(s,x)\right] \nonumber \\
		& = f_F^2(x)e^{J(s,x)}.
	\end{align}
    It follows that
    \begin{align}
	    C_F(s,x)=f_F^2(x)\left[e^{J(s,x)}-1\right]. \label{eq:app_covariance}
    \end{align}

    To evaluate the double spatial integral in eq.~\eqref{eq:app_variance_covariance}, we change the integration variables from $\mathbf{r}_1,\mathbf{r}_2$ to $\mathbf{r}_1,\mathbf{s}$. The allowed integration region for $\mathbf{r}_1$ at a fixed $\mathbf{s}$ is the intersection of two Hubble smoothing spheres of radius $R_H(x)$ separated by $s$. The integration over $\mathbf{r}_1$ is then given by the volume of this intersection $V_{\cap}[R_H(x),s]$ (see eq.~\eqref{eq:overlapV}) since the integrand is independent of $\mathbf{r}_1$. Performing the angular integration for $\mathbf{s}$ and introducing $z=s/R_H(x)$, we obtain
    \begin{align}
	    \sigma_{F,H}^2(x)=\int_0^2 \dd z\,K(z)\,C_F[zR_H(x),x].
    \end{align}
    Here $K(z)$ is defined in eq.~\eqref{eq:kernel}. Substituting eq.~\eqref{eq:app_covariance} into the above expression 
    immediately yields eq.~\eqref{eq:sigmaFH2}.


\bibliographystyle{elsarticle-num}

\bibliography{references}

\setcounter{equation}{0} 
\renewcommand{\theequation}{S\arabic{equation}} 

\titlepage

\setcounter{page}{1}

\begin{center}
	{\LARGE Supplementary Material}
\end{center}

\section{Proof of the propositions on relations between the three completion criteria}\label{sec:appB}

\subsection{Common setup and useful relations} \label{sec:appB1}

We first recall the basic setup adopted in the main text, so that this appendix becomes self-contained. We consider a spatially flat FLRW universe and adopt the Gaussian nucleation model with the vacuum-to-radiation energy-transfer assumption. Natural units $c=\hbar=k_{\rm B}=1$ are adopted, and the bubble wall velocity $v_w=1$ is used throughout without loss of generality. Dots denote derivatives with respect to the cosmic time, while primes denote derivatives with respect to the conformal time. The nucleation rate is parametrized as
\begin{align}
	\Gamma(t)=\Gamma_m\exp\left[-\alpha^2(t-t_m)^2\right]. \label{eq:appBGaussian}
\end{align}
The comoving radius $r(t,t')$ of a bubble nucleated at $t'$\footnote{In this appendix, $t'$ denotes a cosmic time variable rather than the derivative of $t$ with respect to the conformal time $\frac{\dd t}{\dd\eta}$.} and evaluated at $t$ can be computed using the conformal time:
\begin{align}
	\eta(t)=\int_{t_i}^{t}\frac{\dd t'}{a(t')},	\qquad	r(t,t')=\eta(t)-\eta(t'),
	\label{eq:appBcomovingradius}
\end{align}
where $t_i$ is the initial time chosen sufficiently early when nucleation can be neglected, and the scale factor is normalized as $a(t_i)=1$. The false-vacuum fraction (FVF) is given by the JMAK equation
\begin{align}
	f_F(t)=e^{-I(t)},\qquad	I(t)=K\int_{t_i}^{t}\dd t'\,\Gamma(t')a^3(t')[\eta(t)-\eta(t')]^3,\qquad K\equiv\frac{4\pi}{3}.
	\label{eq:appBJMAK}
\end{align}

The cosmological background evolution is governed by the Friedmann equation and the vacuum-to-radiation energy-transfer equation,
\begin{align}
	H^2=\frac{\rho_R+\Delta V_0 f_F}{3M_{\rm Pl}^2}, \qquad \dot{\rho}_R+4H\rho_R=-\Delta V_0\dot f_F, \qquad
	H_V \equiv \sqrt{\frac{\Delta V_0}{3M_{\mathrm{Pl}}^2}}. \label{eq:appBbackground}
\end{align}
Here $\Delta V_0>0$ is the zero-temperature vacuum energy difference, $\rho_R$ is the energy density of the radiation-like sectors, and $H_V$ is the vacuum Hubble scale. We use the dimensionless parameters
\begin{align}
	\lambda_V=\frac{(\Delta V_0)^{1/4}}{M_{\rm Pl}}, \qquad
	\rmbg = \frac{\rho_R^{\rm bg}(t_m)}{\Delta V_0}, \qquad
	q=\frac{\alpha}{H_V}, \qquad
	\gamma=\frac{\Gamma_m}{H_V^4},
\end{align}
where $\rho_R^{\rm bg}(t_m)$ is evaluated in the counterfactual evolution without the phase transition, with the same initial data. We consider the completion properties at fixed $\lambda_V$, with the relevant parameter set given by $\theta=(r_m^{\rm bg},q,\gamma)$.

The comoving radiation-energy budget is defined as
\begin{align}
	\mE(t)\equiv a^4(t)\rho_R(t). \label{eq:appBcreb}
\end{align}
The energy-transfer equation gives
\begin{align}
	\dot\mE=-\Delta V_0a^4\dot f_F. \label{eq:appBdotmE}
\end{align}
Since $\dot f_F\leq 0$, $\mE(t)$ is nondecreasing. The three completion criteria considered in the main text are
\begin{equation}
	\begin{aligned}
		{\rm CMC}:&\quad
		\lim_{t\to\infty}f_F(t)=0,\\
		{\rm PVC}:&\quad
		\lim_{t\to\infty}a^3(t)f_F(t)=0,\\
		{\rm ATC}:&\quad \mE_\infty<\infty.
	\end{aligned}
	\label{eq:appBcriteria}
\end{equation}
One can also derive an energy-transfer equation written in terms of the conformal time
\begin{align}
	\mE'=-\Delta V_0a^4f'_F, \label{eq:appBprimemE}
\end{align}
with the integral form controlling the tail part of the comoving radiation-energy budget
\begin{align}
	\mE_\infty-\mE(\eta) = \Delta V_0\int_\eta^{\eta_\infty} a^4(\xi)[-f'_F(\xi)]\dd\xi. \label{eq:appBintprimeE}
\end{align}
Eqs.~\eqref{eq:appBprimemE}~\eqref{eq:appBintprimeE} are critical in demonstrating the equivalence between PVC and ATC.

The background equations imply two useful identities which read
\begin{align}
	a''=2H_V^2a^3f_F, \label{eq:appBacceleration}
\end{align}
and
\begin{align}
	\dot H = 2H_V^2f_F-2H^2. \label{eq:appBHdot}
\end{align}
In particular, the Hubble rate is bounded from above by its initial value $H_0\equiv H(t_i)>0$ and always decreasing. It is then straightforward to show that
\begin{align}
	H_V\sqrt{f_F}\leq H\leq H_0,\quad a(t)\leq e^{H_0(t-t_i)}. \label{eq:appBbgbound}
\end{align}
Moreover, $a(t)$ never decreases, so that PVC implies CMC.

The evolution equations can be cast into a smooth finite-dimensional system of ordinary differential equations (ODEs). The boundedness of the nucleation rate, Hubble rate, and the scale factor excludes finite-time blow up. Standard ODE theory ensures a unique future-global
classical solution and continuous dependence on parameters over every finite cosmic-time interval. The same continuity holds at any finite conformal time reached by a reference solution.

We next collect the estimates on the nucleation integrals that will be repeatedly used. Define
\begin{align}
	\mathcal W(\eta)\equiv a^4(\eta)\Gamma[t(\eta)], \qquad Q_\Gamma\equiv\int_0^{\eta_\infty}\mathcal W(\xi)\dd\xi =\int_{t_i}^{\infty}a^3(t)\Gamma(t)\dd t.
	\label{eq:appBmW}
\end{align}
The Gaussian tail, together with the upper bound on $a(t)$ in eq.~\eqref{eq:appBbgbound}, guarantees $Q_\Gamma<\infty$.
In terms of conformal time, the JMAK exponent and its first three derivatives are
\begin{equation}
	\begin{aligned}
		I(\eta)
		&=K\int_0^\eta \dd\xi\,
		\mathcal W(\xi)(\eta-\xi)^3,\\
		I'(\eta)
		&=3K\int_0^\eta \dd\xi\,
		\mathcal W(\xi)(\eta-\xi)^2,\\
		I''(\eta)
		&=6K\int_0^\eta \dd\xi\,
		\mathcal W(\xi)(\eta-\xi),\\
		I'''(\eta)
		&=6K\int_0^\eta \dd\xi\,\mathcal W(\xi).
	\end{aligned}
	\label{eq:appBJMAKexponent}
\end{equation}
Consequently,
\begin{equation}
	\begin{aligned}
		0\leq I(\eta)&\leq KQ_\Gamma\eta^3,\\
		0\leq I'(\eta)&\leq3KQ_\Gamma\eta^2,\\
		0\leq I''(\eta)&\leq6KQ_\Gamma\eta,\\
		0\leq I'''(\eta)&\leq6KQ_\Gamma.
	\end{aligned}
	\label{eq:appBIbounds}
\end{equation}

On the other hand, any finite initial interval $[0,\eta_s]$ with $\eta_s>0$ carries a strictly positive nucleation budget,
\begin{align}
	Q_s\equiv\int_0^{\eta_s}\mathcal W(\xi)\dd\xi>0.
\end{align}
Retaining only bubbles nucleated in this interval gives the lower bounds
\begin{equation}
	\begin{aligned}
		I(\eta)&\geq KQ_s(\eta-\eta_s)^3,\\
		I'(\eta)&\geq3KQ_s(\eta-\eta_s)^2,\\
		f_F(\eta)&\leq e^{-KQ_s(\eta-\eta_s)^3},
	\end{aligned}
	\qquad \eta_s\leq\eta<\eta_\infty.
	\label{eq:appBmultiplebounds}
\end{equation}
In particular, for $\eta\geq2\eta_s$,
\begin{align}
	I(\eta)\geq\frac{KQ_s}{8}\eta^3, \qquad I'(\eta)\geq\frac{3KQ_s}{4}\eta^2.
	\label{eq:appBearlybubble}
\end{align}

These estimates also yield a useful characterization of CMC in terms of the total future conformal time,
\begin{align}
	{\rm CMC}\quad\Longleftrightarrow\quad\eta_\infty\equiv\lim_{t\to\infty}\eta(t)=\infty. \label{eq:appBCMCconformal}
\end{align}
Indeed, if $\eta_\infty<\infty$, the upper bound in eq.~\eqref{eq:appBIbounds} implies that $I$ remains finite and hence $f_F$ approaches a strictly positive limit. Conversely, if $\eta_\infty=\infty$, the lower bound in eq.~\eqref{eq:appBmultiplebounds} implies $I\to\infty$ and therefore $f_F\to0$.

\subsection{Equivalence between PVC and ATC} \label{sec:appB2}

The equivalence between PVC and ATC can be most easily proved using conformal time. First, we show ${\rm ATC}\Rightarrow{\rm PVC}$. Suppose ATC is valid, we are then able to exclude a finite $\eta_\infty$. Otherwise, let us examine eq.~\eqref{eq:appBprimemE}. Using eq.~\eqref{eq:appBCMCconformal}, a finite $\eta_\infty$ implies the failure of CMC and thus $f_F\geq f_F(\eta_\infty)>0$. On the other hand, $I'$ also has a positive lower bound from eq.~\eqref{eq:appBearlybubble}. Therefore we have $-f'_F=I'f_F\geq c_f>0$, where $c_f$ is some time-independent positive number. As to the scale factor, using eq.~\eqref{eq:appBcomovingradius} and $H\leq H_0$ one can show that $a(\eta)\geq\frac{1}{H_0(\eta_\infty-\eta)}$. Combining with $-f'_F\geq c_f>0$, we then obtain the following lower bound on $\mE'$:
\begin{align}
	\mE'\geq\frac{\Delta V_0 c_f}{H_0^4(\eta_\infty-\eta)^4}. \label{eq:appBmEplb}
\end{align}
A finite $\eta_\infty$ would imply the integral of the right-hand side of eq.~\eqref{eq:appBmEplb} diverges, in contradiction to the requirement that ATC succeeds. We therefore conclude that successful ATC implies $\eta_\infty=\infty$, and thus CMC succeeds with $f_F(\infty)=0$. Eq.~\eqref{eq:appBintprimeE} can then be employed to yield an inequality for the tail part of the comoving radiation-energy budget
\begin{align}
	\mE_\infty-\mE(\eta)\geq\Delta V_0 a^4(\eta)f_F(\eta).
\end{align}
The success of ATC means $\mE_\infty-\mE(\eta)\rightarrow 0$ as $\eta\rightarrow\infty$, which immediately leads to
\begin{align}
	\lim_{\eta\to\infty}a^4 f_F = 0.
\end{align}
Since the scale factor never decreases, we arrive at PVC.

Next, we show ${\rm PVC}\Rightarrow{\rm ATC}$. Assume PVC holds, eq.~\eqref{eq:appBacceleration} then implies $a''=2H_V^2a^3f_F\rightarrow 0$ as $\eta\rightarrow\infty$, from which we infer
\begin{align}
	a(\eta)=\mathcal{O}(\eta^2),\qquad \eta\rightarrow\infty. \label{eq:appBaeta}
\end{align}
On the other hand, eqs.~\eqref{eq:appBIbounds}~\eqref{eq:appBmultiplebounds} tell us that $I'$ and $f_F$ are bounded from above:
\begin{align}
	I'(\eta)\leq3KQ_\Gamma\eta^2,\qquad f_F(\eta)&\leq e^{-KQ_s(\eta-\eta_s)^3}. \label{eq:appBIpfF}
\end{align}
The use of eqs.~\eqref{eq:appBaeta}~\eqref{eq:appBIpfF} in eq.~\eqref{eq:appBprimemE} yields an upper bound on $\mE'$:
\begin{align}
	\mE'\leq C\eta^{10} e^{-KQ_s(\eta-\eta_s)^3}, \text{for sufficiently large}\,\eta. \label{eq:appBmEpbound}
\end{align}
Since the integral of the right-hand side of eq.~\eqref{eq:appBmEpbound} converges, we conclude that the integral of $\mE'$ must also converge, i.e. ATC also succeeds.

Therefore, PVC and ATC are completely equivalent for the model studied in this work.

\subsection{From CMC to PVC} \label{sec:appB3}

In this subsection, we prove that success of CMC with a dimensionless nucleation amplitude $\gamma_1$ implies success of PVC with any dimensionless nucleation amplitude $\gamma_2>\gamma_1$, when other parameters $(\rmbg,q,v_w)$ are kept fixed. The main idea is to control the growth of both $f_F$ and $a$ as a function of the cosmic time $t$, so that $a^3 f_F$ can be bounded by a function that decreases sufficiently fast as $t\rightarrow\infty$.

First, we show that, if CMC holds, then for each $\varepsilon >0$, there exists a finite $T_\varepsilon >0$, such that
\begin{align}
	H_V^2 t^2 f_F(t)\leq 1+\varepsilon,\qquad \text{for all}\qquad t\geq T_\varepsilon. \label{eq:appBfFdecay}
\end{align}
To this end, we introduce several auxiliary functions
\begin{align}
	F(\eta)\equiv e^{-I(\eta)/2},\qquad p(\eta)\equiv\frac{I'(\eta)}{2},\qquad \omega(\eta)\equiv\frac{p'(\eta)}{p(\eta)^2}.
\end{align}
and
\begin{align}
	J(\eta)\equiv\int_\eta^\infty F(\xi)\dd\xi,\qquad g(\eta)\equiv\frac{F(\eta)}{p(\eta)}.
\end{align}
For each $\epsilon >0$, $0\leq\omega(\eta)\leq\epsilon$ can be satisfied for some sufficiently large $\eta$. Together with the fact that $-g'=F(1+\omega)$, this implies that for each $\epsilon >0$, there exists some $\eta_\epsilon>0$ such that
\begin{align}
	J(\eta)\geq\frac{F(\eta)}{(1+\epsilon)p(\eta)},\qquad \eta\geq\eta_\epsilon. \label{eq:appBJeta1}
\end{align}

The integration of $(1/a)'=-H$, together with the first inequality in eq.~\eqref{eq:appBbgbound}, leads to the following key inequality
\begin{align}
	a(\eta)\leq\frac{1}{H_V J(\eta)}. \label{eq:appBJeta2}
\end{align}
Thus a lower bound on $H$ gives rise to an upper bound on $a$. This is counter-intuitive but reasonable, because the upper bound on $a$ is using the conformal time $\eta$ instead of the cosmic time $t$ as an argument. The relation $(1/a)'=-H$ which employs the derivative with respect to $\eta$ implies that a too large initial scale factor would be incompatible with a lower bound on $H$. By combining eq.~\eqref{eq:appBJeta1} and eq.~\eqref{eq:appBJeta2}, the following inequality can be derived using $\frac{\dd t}{\dd\eta}=a$,
\begin{align}
	\frac{\dd t}{\dd\eta}\leq\frac{1+\epsilon}{H_V}\left(\frac{1}{F}\right)',
\end{align}
for each $\epsilon>0$ and all $\eta>\eta_\epsilon$. Upon integration, we obtain for some finite positive constant $C_\epsilon$
\begin{align}
	H_V t(\eta)F(\eta)\leq 1+\epsilon+H_V C_\epsilon F(\eta).
\end{align}
At large $\eta$, $F(\eta)$ decreases exponentially, so the term $H_V C_\epsilon F(\eta)$ can also be bounded by $\epsilon$ for sufficiently large $\eta$. Assuming this is satisfied, we may then choose $\epsilon=\frac{\sqrt{1+\varepsilon}-1}{2}$ so that eq.~\eqref{eq:appBfFdecay} is satisfied.

Next, we prove the following result concerning the solutions of phase-transition dynamics and cosmological evolution of two parameter points having different dimensionless nucleation amplitudes $\gamma_1$ and $\gamma_2$ but the same set of remaining parameters. Let $r=\frac{\gamma_2}{\gamma_1}>1$ be the ratio of two nucleation amplitudes, and we use subscripts $1,2$ to label the corresponding solutions, then at any common cosmic time $t>0$
\begin{align}
	H_2(t)<H_1(t),\qquad a_2(t)<a_1(t),\qquad I_2(t)\geq I_1(t),\qquad f_{F,2}(t)\leq f_{F,1}(t)^r<f_{F,1}(t). \label{eq:appBcompare}
\end{align}
The proof utilizes the following kernel function which appears in the expression for $I$:
\begin{align}
	L_H(t,s) = a(s)[\eta(t)-\eta(s)] = \int_s^t \exp\left[-\int_s^u H(v)\,\dd v\right]\dd u. \label{eq:appBLH}
\end{align}
The JMAK exponent is then $I(t) = K\gamma \int_0^t G(s)L_H(t,s)^3\,\dd s$ with $G(s)=H_V^4\exp\left[-\alpha^2(s-t_m)^2\right]$ being the Gaussian kernel, and we have set $t_i=0$ for simplicity. Since $L_H(t,s)$ can be expressed in terms of integration of the Hubble history, as shown in eq.~\eqref{eq:appBLH}, if $H_2\leq H_1$ is satisfied on some finite time interval $[0,T]$, we immediately obtain on the same interval,
\begin{align}
	L_{H_2}(t,s)\geq L_{H_1}(t,s),\qquad I_2(t)\geq r I_1(t),\qquad f_{F,2}(t)\leq f_{F,1}(t)^r. \label{eq:appBcompare1}
\end{align}

We now establish the initial ordering using bounds on background quantities. $0<H_i\leq H_0$ leads to
\begin{align}
	e^{-H_0 t}(t-s)\leq L_{H_i}(t,s)\leq t-s.
\end{align}
Introducing an auxiliary function $W(t)= K\int_0^t G(s)(t-s)^3\,\dd s$, for $t>0$ we then have
\begin{align}
	\gamma_i e^{-3H_0 t}W(t)\leq I_i(t)\leq\gamma_i W(t). \label{eq:appBwindow}
\end{align}
Since two nucleation amplitudes are different, it is always possible to find a short initial time interval $(0,t_{\rm in}]$ on which the two windows defined by eq.~\eqref{eq:appBwindow} associated with $i=1,2$ do not overlap, and thus $I_2>I_1, f_{F,2}<f_{F,1}$ on this interval. We then define $D_H=H_2-H_1$. From eq.~\eqref{eq:appBHdot} the following equation for $D_H$ can be derived:
\begin{align}
	\dot{D}_H + 2(H_1 + H_2)D_H = 2H_V^2(f_{F,2} - f_{F,1}), \qquad D_H(0) = 0. \label{eq:appBDHdot}
\end{align}
Its solution can be written as
\begin{align}
	D_H(t) = \frac{2H_V^2}{\mu(t)} \int_0^t \mu(s) [f_{F,2}(s) - f_{F,1}(s)]\,\dd s,\qquad \mu(t) = \exp\left[2\int_0^t (H_1(s)+H_2(s))\,\dd s\right].
\end{align}
Since $\mu(t)$ is positive, $f_{F,2}<f_{F,1}$ on $(0,t_{\rm in}]$ then implies $D_H<0$, that is $H_2<H_1$ on this initial interval. In fact, the ordering $H_2<H_1$ persists for all $t>0$, otherwise there would exist some $t_*>0$ such that $H_2(t_*)=H_1(t_*)$, but before $t_*$ we have $H_2<H_1$. However, such a situation would be inconsistent with eq.~\eqref{eq:appBDHdot} evaluated at $t_*$. The contradiction implies that $t_*$ cannot exist. Thus we have $H_2<H_1$ for all $t>0$, and the rest of the proof is straightforward by arguments around eq.~\eqref{eq:appBcompare1}.

Finally, we return to the analysis of $a^3 f_F$ in order to prove the statement claimed at the beginning of this subsection. Now, for the two solutions corresponding to two different dimensionless nucleation amplitudes $\gamma_2>\gamma_1$, eqs.~\eqref{eq:appBfFdecay}~\eqref{eq:appBcompare} give
\begin{align}
	f_{F,1}(t) &\leq \frac{2}{(H_V t)^2}, \qquad
	f_{F,2}(t) \leq f_{F,1}(t)^r \leq \frac{2^r}{(H_V t)^{2r}}. \label{eq:appBaux1}
\end{align}
For the solution corresponding to $\gamma_2$, we choose some $T_0>0$, and introduce
\begin{align}
	\tau = \ln(t/T_0), \qquad Y(\tau) = tH_2(t), \qquad t = T_0 e^\tau.
\end{align}
Using eq.~\eqref{eq:appBHdot}, one can show that $Y(\tau)$ satisfies
\begin{align}
	\frac{\dd Y}{\dd\tau} = Y-2Y^2+\Psi(T_0 e^\tau),\qquad \Psi(t)\equiv 2H_V^2 t^2 f_{F,2}(t). \label{eq:appBYtau}
\end{align}
Note that since $r>1$, $\Psi(t)\leq \frac{2^{r+1}}{(H_V t)^{2r-2}}\rightarrow 0$ as $t\rightarrow\infty$. This means the $\Psi(T_0 e^\tau)$ term can be arbitrarily small for sufficiently large $\tau$. It is then straightforward to show that $Y(\tau)\leq 2/3$ for sufficiently large $\tau$, or equivalently
\begin{align}
	H_2(t)\leq\frac{2}{3t},\qquad t\geq T,
\end{align}
for some large $T\geq T_0$. This in turn yields $a_2(t)^3\leq Z_1 t^2$ for sufficiently large $t$, with $Z_1$ being some positive constant. Combining with eq.~\eqref{eq:appBaux1} we obtain $a_2^3(t) f_{F,2}(t)\leq Z_2 t^{2-2r}\rightarrow 0$ as $t\rightarrow\infty$, and thus PVC succeeds for $\gamma_2$.

\subsection{Common completion boundary}
\label{sec:appB4}

Let $\Theta$ denote the allowed parameter space of $\theta=(r_m^{\mathrm{bg}},q,\gamma)$
introduced in~\ref{sec:appB1}, equipped with the topology inherited from $\mathbb{R}^3$. All parameter neighborhoods are restricted to $\Theta$. In particular, at every parameter point, $\gamma$ can be increased or decreased by a sufficiently small amount while keeping $r_m^{\mathrm{bg}}$ and $q$ fixed.

A set is relatively open if each of its points has a parameter neighborhood within $\Theta$ contained in that set. A point
belongs to its relative boundary if every such neighborhood	intersects both the set and its complement in $\Theta$.
We write $\partial_\Theta S$, $\overline{S}$ and $\operatorname{int}_\Theta S$ for the boundary, closure and
interior of a set $S$, all taken relative to $\Theta$.

Define
\begin{align}
	S_M &\equiv \{\theta\in\Theta:\text{CMC holds}\},
	&
	S_P &\equiv \{\theta\in\Theta:\text{PVC holds}\},
	\\
	S_F &\equiv \Theta\setminus S_M,
	&
	S_C &\equiv S_M\setminus S_P,
\end{align}
where each completion criterion is evaluated on the	self-consistent solution corresponding to $\theta$. Since PVC implies CMC, $S_P\subseteq S_M$, and $\Theta=S_P\cup S_C\cup S_F$ is a partition into three pairwise disjoint sets.

In this subsection, we sketch the proof of the following statement. Under the assumptions of~\ref{sec:appB1}, the sets $S_P$ and $S_F$ are relatively open, and
\begin{align}
	S_C
	=\partial_\Theta S_P
	=\partial_\Theta S_M
	=\partial_\Theta S_F. \label{eq:appBT1}
\end{align}
Furthermore,
\begin{align}
	\overline{S_P}=S_M,
	\qquad
	\operatorname{int}_\Theta S_M=S_P. \label{eq:appBT2}
\end{align}
Consequently, $S_M$ and $S_C$ are relatively closed, and $S_C$ has empty relative interior. CMC and PVC are therefore equivalent away from their common boundary, whereas CMC	holds and PVC fails at every point on that boundary. By the equivalence established in~\ref{sec:appB2}, ATC has the same completion boundary and also fails there.

Parameter points at which CMC holds but PVC fails must belong to the boundary of CMC and PVC. This can be easily demonstrated using $\text{CMC}(\gamma_1)\Rightarrow\text{PVC}(\gamma_2)$ when $\gamma_2>\gamma_1$, proved in~\ref{sec:appB3}. The upshot is
\begin{align}
	S_C\subseteq \partial_\Theta S_P\cap\partial_\Theta S_M\cap\partial_\Theta S_F.
\end{align}

We then prove $S_F$ is relatively open, that is, for each $\theta_*\in S_F$, there exists a relative neighborhood $U$ in which all parameters correspond to CMC failure. The key observation is that CMC is an asymptotic property, while the ODE continuity properties with respect to parameters are valid on finite time intervals. Therefore, the crucial step is to find a criterion which can be established on finite time intervals but can guarantee CMC failure in the infinite future.

To establish such a criterion, we introduce an auxiliary conditional bounding function $B(\theta,t_0)$ as follows ($\theta$ denotes the parameter point $(\rmbg,q,\gamma)$). Choose a finite time $t_0$ with
\begin{align}
	a_0=a(t_0),\qquad f_0=f_F(t_0)>0,\qquad J_k \equiv \left.\frac{\dd^k I}{\dd\eta^k}\right|_{\eta=\eta(t_0)},\,k=1,2,3.
\end{align}
We then introduce
\begin{align}
	h = H_V \sqrt{\frac{f_0}{2}}, \qquad
	d = \frac{1}{a_0 h}, \qquad
	R_{\Gamma}(\theta,t_0) = \int_{t_0}^{\infty} \Gamma(s)\,\dd s.
\end{align}
The dimensionless function $B(\theta,t_0)$ is then defined as
\begin{align}
	B(\theta,t_0) = J_1 d + \frac{J_2 d^2}{2} + \frac{J_3 d^3}{6} + \frac{K}{h^3} R_{\Gamma}(\theta,t_0).
\end{align}
One can show that, if $B(\theta,t_0)<\ln 2$, then for all $t\geq t_0$ we have $f_F(t)>f_0/2$, and thus CMC fails. The first step is to prove on a finite interval $[t_0,T]$ that if $f_F(u)\geq f_0/2$ for $u\in [t_0,T]$, then $I(t)-I(t_0)\leq B(\theta,t_0)$ for all $t\in [t_0,T]$. The idea is to split $I(t)$ (and its change $\Delta I\equiv I(t)-I(t_0)$) into contributions from old bubbles nucleated before $t_0$, and new bubbles nucleated after $t_0$. The old-bubble contribution satisfies the exact identity
\begin{align}
	\Delta I_{\rm old}(t;t_0)=J_1 \Delta_\eta(t) + \frac{1}{2}J_2[\Delta_\eta(t)]^2 + \frac{1}{6}J_3[\Delta_\eta(t)]^3,
\end{align}
where $\Delta_\eta(t)\equiv\eta(t)-\eta(t_0)$. Using $H\geq h$ on the interval $[t_0,T]$ we obtain $\Delta_\eta<d$, and thus
\begin{align}
	\Delta I_{\rm old}(t;t_0)\leq  J_1 d + \frac{J_2 d^2}{2} + \frac{J_3 d^3}{6},\,\text{for all}\,t\in [t_0,T].
\end{align}
As to the new-bubble contribution, using $H(v)\geq h$ we may obtain $L_H(t,s)\leq 1/h$, where $L_H(t,s)$ is defined in eq.~\eqref{eq:appBLH}. The new-bubble contribution then satisfies $I_{\rm new}\leq \frac{K}{h^3} R_{\Gamma}$. Combining both the old-bubble and new-bubble contributions, we then obtain $I(t)-I(t_0)\leq B(\theta,t_0)$ for the finite interval $[t_0,T]$ under consideration.

Now let us assume $B(\theta,t_0)<\ln 2$, and suppose that there exists some $t_*>t_0$ such that $f_F(t_*)=f_0/2$ but for $t_0\leq t\leq t_*$ we have $f_F(t)\geq f_0/2$. We may then apply the finite-interval argument above to obtain
\begin{align}
	I(t_*)-I(t_0)\leq B(\theta,t_0)<\ln 2. \label{eq:appBIbound}
\end{align}
However, $f_F(t_*)=f_0/2$ implies $I(t_*)-I(t_0)=\ln 2$, which contradicts eq.~\eqref{eq:appBIbound}. Therefore, the time $t_*$ satisfying the above condition does not exist, which in turn implies $f_F(t)>f_0/2$ for all $t\geq t_0$.

We now pick a $\theta_*\in S_F$. When we increase $t_0$ at this parameter point, $d\rightarrow 0$, so that $B(\theta_*,t_0)\rightarrow 0$. Therefore, there must exist some $t_0$ such that $B(\theta_*,t_0)<\ln 2$ can be satisfied. From now on we keep this $t_0$ fixed when we vary parameters. Then, since the function $B(\theta,t_0)$ depends on the parameters in $\theta$ in a continuous manner, there must exist some neighborhood of $\theta_*$ such that in this neighborhood, $B(\theta,t_0)<\ln 2$ still holds, and thus CMC fails for all parameters in this neighborhood. This concludes the proof that $S_F$ is relatively open.

Next, we prove $S_P$ is relatively open. As a preparation, we employ the following strategy to bound the scale factor. First we pick a solution $a(\eta)$ and choose a finite conformal time $\eta_0$ within its reach. Suppose there exists some constant $C>0$ and a known non-negative function $E(\eta)$ (not to be confused with the comoving radiation-energy budget or its dimensionless version) such that the solution $a(\eta)$ satisfies
\begin{align}
	a''(\eta)\leq Ca(\eta)^3 E(\eta),\qquad \eta_0\leq\eta<\eta_\infty. \label{eq:appBappbound}
\end{align}
If one can further construct some function $P(\eta)$ such that
\begin{align}
	P''(\eta)>CP(\eta)^3 E(\eta),\qquad \eta\geq\eta_0, \label{eq:appBPppbound}
\end{align}
while at the starting point
\begin{align}
	a(\eta_0)<P(\eta_0),\qquad a'(\eta_0)<P'(\eta_0), \label{eq:appBspbound}
\end{align}
then it is straightforward to show that $a<P$ always holds within the actual domain of both functions.

The general idea now is to consider a fiducial parameter point $\theta_*\in S_P$, and show that there exists an open neighborhood $U$ of $\theta_*$ such that PVC holds for all $\theta\in U$. First, for a fixed $\theta_*\in S_P$, PVC and eq.~\eqref{eq:appBacceleration} imply ($V_F\equiv a^3f_F$)
\begin{align}
	a_{\theta_*}''(\eta) = 2H_V(\theta_*)^2 V_{F,\theta_*}(\eta) \longrightarrow 0
	\qquad (\eta \to \infty).
\end{align}
Integration then leads to $a_{\theta_*}(\eta)\leq C_a\eta^2$ for sufficiently large $\eta$. We then choose an early conformal time $\eta_s$ and use the bounds in eq.~\eqref{eq:appBmultiplebounds} to obtain for sufficiently large $\eta$
\begin{align}
	0 \leq a_{\theta_*}''(\eta) \leq 2H_V(\theta_*)^2 C_a^3 \eta^6 e^{-KQ_s(\theta_*)(\eta-\eta_s)^3}.
\end{align}
The integral of the right-hand side converges. Therefore we may define
\begin{align}
	b_* = H_0(\theta_*) + \int_0^\infty a_{\theta_*}''(\eta)\,\dd\eta,
\end{align}
and obtain the following limits
\begin{align}
	\lim_{\eta\to\infty} a_{\theta_*}'(\eta) = b_* > 0,
	\qquad
	\lim_{\eta\to\infty} \frac{a_{\theta_*}(\eta)}{\eta} = b_*. \label{eq:appBalimits}
\end{align}

At this moment, let us fix the value of $\eta_s$ used in the fiducial case, and consider a relatively open neighborhood $U_1$ of $\theta_*$, so that each solution in $U_1$ can reach $\eta_s$. Due to continuity, one may choose positive constants $A_*,C_*$ so that
\begin{align}
	K Q_s(\theta) \geq A_*, \qquad
	2H_V(\theta)^2 \leq C_*
	\qquad (\theta \in U_1).
\end{align}
Now we define the common $E(\eta)$ function
\begin{align}
	E_{\mathrm{op}}(\eta) = e^{-A_*(\eta-\eta_s)^3}.
\end{align}
The following inequalities can then be obtained by using background identities and eq.~\eqref{eq:appBacceleration}
\begin{align}
	f_{F,\theta}(\eta) \leq E_{\mathrm{op}}(\eta), \qquad a_\theta''(\eta) \leq C_* a_\theta(\eta)^3 E_{\mathrm{op}}(\eta), \qquad \eta_s \leq \eta < \eta_\infty(\theta). \label{eq:appBinequalities}
\end{align}
The first inequality, $f_{F,\theta}(\eta) \leq E_{\mathrm{op}}(\eta)$, will be reserved for bounding $V_F$ in order to prove PVC. The second inequality in eq.~\eqref{eq:appBinequalities} corresponds to the condition eq.~\eqref{eq:appBappbound} required to bound the scale factor. As to the auxiliary bounding function $P(\eta)$, we consider the following simple ansatz
\begin{align}
	P_{\mathrm{op}}(\eta) = \frac{D}{\eta_0}\eta^2 \qquad (\eta \geq \eta_0).
\end{align}
Here $D>b_*$ and $\eta_0>\eta_s$ are constants chosen such that eqs.~\eqref{eq:appBPppbound} and~\eqref{eq:appBspbound} can be satisfied, that is
\begin{align}
	P_{\mathrm{op}}''(\eta) > C_* P_{\mathrm{op}}(\eta)^3 E_{\mathrm{op}}(\eta) \qquad (\eta \geq \eta_0), \label{eq:appBP2}
\end{align}
and
\begin{align}
	a_{\theta_*}(\eta_0) < D\eta_0, \qquad a_{\theta_*}'(\eta_0) < 2D. \label{eq:appBsp2}
\end{align}
Eq.~\eqref{eq:appBalimits} guarantees that eq.~\eqref{eq:appBsp2} can be satisfied at late times for $D>b_*$, while eq.~\eqref{eq:appBP2} can be satisfied if
\begin{align}
	A_*\eta_0(\eta_0-\eta_s)^2 \geq 2, \qquad \frac{C_*D^2}{2}\eta_0^4 E_{\mathrm{op}}(\eta_0) < 1.
\end{align}
The ODE continuity with respect to $\theta$ allows the generalization of eq.~\eqref{eq:appBsp2} to some open neighborhood $U\subset U_1$ of $\theta_*$, so that
\begin{align}
	a_\theta(\eta_0) < P_{\mathrm{op}}(\eta_0), \qquad a_\theta'(\eta_0) < P_{\mathrm{op}}'(\eta_0)	\qquad (\theta \in U).
\end{align}
Now let us fix $U$ and an arbitrary parameter point $\theta\in U$, the bounding strategy for the scale factor based on the conditions eqs.~\eqref{eq:appBappbound}~\eqref{eq:appBPppbound}~\eqref{eq:appBspbound} then implies
\begin{align}
	a_\theta(\eta) < P_{\mathrm{op}}(\eta) \qquad (\eta_0 \leq \eta < \eta_\infty(\theta)).
\end{align}
One can further exclude the possibility that $\eta_\infty(\theta)$ is finite. This is because $a(\eta)$ is now bounded by $P(\eta)$. Since $\dd t/\dd\eta=a$, finite $\eta_\infty(\theta)$ would imply that $t$ cannot reach infinity, in contradiction to the existence of a future-global solution guaranteed by the standard ODE theory. Finally, we have
\begin{align}
	0 < V_{F,\theta}(\eta) \leq P_{\mathrm{op}}(\eta)^3 E_{\mathrm{op}}(\eta) = \frac{D^3}{\eta_0^3}\eta^6 e^{-A_*(\eta-\eta_s)^3} \longrightarrow 0.
\end{align}
Therefore, for this fixed parameter point $\theta\in U$ PVC also holds. We therefore conclude that $S_P$ is a relatively open set.

Using the relative openness of $S_F$ and $S_P$, together with $\text{PVC}\Rightarrow\text{CMC}$ and $\text{CMC}(\gamma_1)\Rightarrow\text{PVC}(\gamma_2)$ when $\gamma_2>\gamma_1$, one may then obtain eqs.~\eqref{eq:appBT1}~\eqref{eq:appBT2} via simple topological arguments.

\end{document}